\documentclass{egpubl}

\JournalSubmission    
 \electronicVersion 

\ifpdf \usepackage[pdftex]{graphicx} \pdfcompresslevel=9
\else \usepackage[dvips]{graphicx} \fi

\PrintedOrElectronic

\usepackage{t1enc,dfadobe}

\usepackage{egweblnk}
\usepackage{cite}
\usepackage{amsmath}
\usepackage{mathtools}
\usepackage{amssymb}
\usepackage[normalem]{ulem}
\usepackage{lineno}
\usepackage{siunitx}
\usepackage{bbm}

\title{RealMat: Realistic Materials with Diffusion and \\Reinforcement Learning}

\author[Zhou et al.]
{\parbox{\textwidth}{\centering X. Zhou$^{1,2,3}$, P. Figueiredo$^{1}$, M. Ha\v san$^{3}$, V. Deschaintre$^{3}$, P. Guerrero$^{3}$, Y. Hu$^{3}$, N. Khademi Kalantari$^{1}$
        }
        \\
{\parbox{\textwidth}{\centering $^1$Texas A\&M University, 
        $^2$Max Planck Institute for Informatics, 
      $^3$Adobe Research
       }
}
}

\begin{document}
\newcommand{\method}{RealMat}
\newcommand{\mlp}{\textbf{M}}
\newcommand{\gen}{\textbf{G}}
\newcommand{\decoder}{\textbf{E}}
\newcommand{\loss}{\mathcal{L}_E}

\newcommand{\relit}{{\textbf{I}_{\tiny nr}}}
\newcommand{\real}{{\textbf{I}_{\tiny real}}}
\newcommand{\realcrop}{{\textbf{I}_{\tiny crop}}}
\newcommand{\rendered}{{\textbf{I}_{\tiny ar}}}

\newcommand{\vde}[1]{\textcolor{orange}{{VDE: #1}}}
\newcommand{\xz}[1]{\textcolor{black}{{#1}}} 
\newcommand{\mh}[1]{\textcolor{blue}{{MH:#1}}}
\newcommand{\paul}[1]{\textcolor{brown}{{PG: #1}}}
\newcommand{\ks}[1]{\textcolor{magenta}{{KS: #1}}}
\newcommand{\hyw}[1]{\textcolor{violet}{{YH: #1}}}


\newenvironment{tight_itemize}{
\begin{itemize}[leftmargin=10pt,nosep]
  \setlength{\topsep}{0pt}
  \setlength{\itemsep}{0pt}
  \setlength{\parskip}{0pt}
  \setlength{\parsep}{0pt}
}{\end{itemize}}

\teaser{
 \centering
 \includegraphics[width=\linewidth]{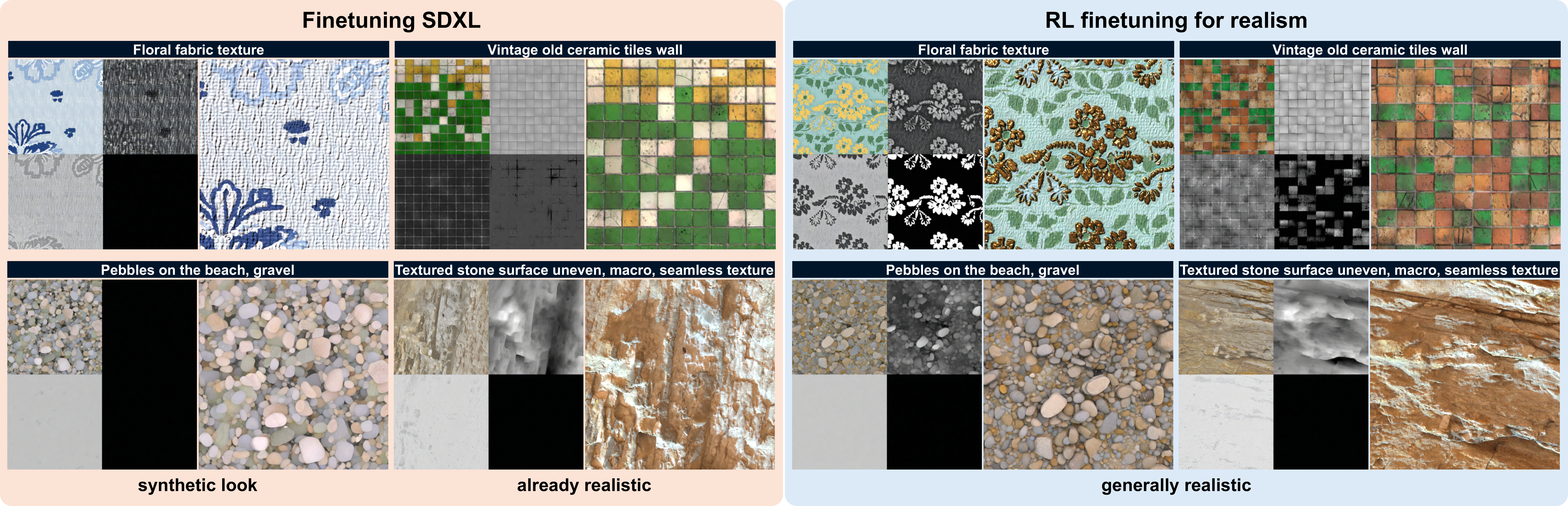}
 \caption{We propose \method{}, a diffusion-based realistic text-to-material generator. We first finetune a Stable Diffusion XL (SDXL) model pretrained on images to generate detailed material maps using synthetic training data (left). Although high-quality materials are generated in many cases, this finetuning shifts the distribution of SDXL toward a more synthetic appearance. To address this, we further finetune our model using reinforcement learning (RL) with a realism reward function (right). We show samples and maps (in clockwise order from top left: albedo, height, metallicity, and roughness) with fixed text prompts and seeds. With RL, the generated material distribution shifts towards more realistic overall.
 } 
	\label{fig:teaser}
}

\maketitle

\makeatletter
\def\ps@arxiv{%
  \def\@oddhead{}%
  \def\@evenhead{}%
  \def\@oddfoot{\hfil\thepage\hfil}%
  \def\@evenfoot{\hfil\thepage\hfil}%
}
\makeatother

\pagestyle{arxiv}
\thispagestyle{arxiv}

\begin{abstract}

Generative models for high-quality materials are particularly desirable to make 3D content authoring more accessible. However, the majority of material generation methods are trained on synthetic data. Synthetic data provides precise supervision for material maps, which is convenient but also tends to create a significant visual gap with real-world materials. Alternatively, recent work used a small dataset of real flash photographs to guarantee realism, however such data is limited in scale and diversity. To address these limitations, we propose \method{}, a diffusion-based material generator that leverages realistic priors, including a text-to-image model and a dataset of realistic material photos under natural lighting. In \method{}, we first finetune a pretrained Stable Diffusion XL (SDXL) with synthetic material maps arranged in $2 \times 2$ grids. This way, our model inherits some realism of SDXL while learning the data distribution of the synthetic material grids. Still, this creates a realism gap, with some generated materials appearing synthetic. We propose to further finetune our model through reinforcement learning (RL), encouraging the generation of realistic materials. We develop a realism reward function for any material image under natural lighting, by collecting a large-scale dataset of realistic material images. We show that this approach increases generated materials' realism compared to our base model and related work.




\end{abstract}  

\section{Introduction}

Material reflectance properties play a key role in the appearance of a 3D scene. Material design by artists traditionally required manual work by experts and specialized software tools. More recently, generative techniques have been proposed~\cite{Guo2020, guerrero2022matformer, zhou2022tilegen, vecchio2024matfuse, xue2024reflectancefusion}. However, these still fall short of achieving sufficient realism, since they are trained on synthetic data. Material generation exclusively from real photos has also been explored~\cite{zhou2023photomat}, but this approach still suffers from limited real training data. \xz{For example, the largest measured material dataset~\cite{ma2023opensvbrdf} contains only slightly over 1,000 samples, which can cause overfitting issues for diffusion-based generation models. To improve realism of material generators,} we propose \method{}, which leverages the priors learned by a text-to-image diffusion model (we use Stable Diffusion XL~\cite{rombach2022high, podell2023sdxl}), combined with a reinforcement learning (RL) approach~\cite{ddpo} to further improve the realism of the generated materials.

Finetuning diffusion models has been proposed for various tasks, showing that the priors from the original models can usually be preserved while adapting the model to new domains. However, these tasks typically still target RGB images, while \xz{Spatially Varying Bidirectional Reflectance Distribution Functions (SVBRDFs)} require multiple channels (typically diffuse albedo, heights/normals, roughness, metallicity, etc.).
To adapt SDXL to generate materials rather than RGB images, we propose to organize the albedo, height map, roughness and metallicity into a $2 \times 2$ grid within a single RGB image. Using this approach, we can finetune SDXL using synthetic material data, while retaining the diffusion model prior. This allows our model to generate a much wider range of appearances with higher realism than SVBRDFs generators trained from scratch~\cite{vecchio2024matfuse}. The finetuning process on synthetic data nonetheless pushes the model towards the generation of somewhat more synthetic appearances. To counteract this, we propose to use a reinforcement learning approach inspired by DDPO~\cite{ddpo} to further finetune our model towards generating materials with higher realism. 
Using RL allows us to apply a realism reward directly to the output of the multi-step inference process of a diffusion model. This approach better reflects the quality observed at inference time than traditional training methods, where a loss is applied only to the coarse output of a single denoising step.
We design a reward function that assigns higher rewards to materials that appear more realistic. Specifically, we train a linear layer that takes CLIP image features~\cite{radford2021learning} of the material image under natural lighting, and outputs a realism score. The reward model is trained on a newly filtered and labeled dataset of material images.

Combining the finetuning stage using our material grid approach and the RL stage with a realism reward, \method{} learns to generate photorealistic and diverse materials\xz{, as shown in Fig.~\ref{fig:teaser}}. We evaluate our approach with qualitative comparisons and a user study. 
In summary, we make the following contributions: 
%
\begin{itemize}
\item We propose a simple yet effective strategy to finetune a pretrained text-to-image diffusion model to the SVBRDF map generation task, using a $2 \times 2$ grid approach. 

\item We develop a reward function that can accurately evaluate the realism of rendered materials under natural lighting, and use it in a reinforcement learning (RL) stage to enhance the realism of the final model. 
\end{itemize}

\section{Related Work}

\paragraph*{Image Generation.} Image generation has been a long-standing task in computer vision and computer graphics. Variational Autoencoders (VAEs)~\cite{kingma2014} and Generative Adversarial Networks (GANs)~\cite{GAN} have been widely used for image generation applications. However, VAEs often generate blurry images due to the difficulty of modeling complex joint distribution in image space. GANs can produce sharp images, but they face many challenges like training instability and mode collapse. In recent years, Diffusion Models (DM)~\cite{ho2020denoising} have stood out in image generation tasks due to stable training and high-quality results. With a well-trained noise estimation model, DMs progressively transform input noises into high-quality images through a multi-step denoising process. Built upon DMs, Latent Diffusion Models (LDMs) ~\cite{rombach2022high} were proposed to perform denoising steps in latent space, significantly reducing computational and memory cost for high-resolution ($512 \times 512$ or $768 \times 768$) image generation. To further lift LDMs to higher resolution, Stable Diffusion XL (SDXL)~\cite{podell2023sdxl} utilizes a larger UNet and a second text encoder. Leveraging the priors learned from a very large image dataset~\cite{schuhmann2022laion} SDXL can generate $1\mathrm{k} \times 1\mathrm{k}$ high-fidelity and photo-realistic RGB images. Therefore, we build \method{} on top of SDXL to leverage its strengths, including strong generalization, realism, and high-resolution generation.

\paragraph*{Material Acquisition.} Material reflectance property is a key factor determining the appearance of virtual scenes. Obtaining high-quality photo-realistic materials have been a challenging task in computer graphics industry. Traditional acquisition approaches rely on bulky and expensive hardware such as gonioreflectometer~\cite{Matusik2003, Guarnera2016} to extensively sample the light and view directions to extract 
SVBRDFs from real material samples. Recently, with the development of deep learning techniques, many learning-based methods have been proposed for material acquisition. These methods train material priors using synthetic or real dataset to extract SVBRDFs through either feed-forward process~\cite{Li2017, Li2018, Deschaintre2018, Deschaintre2019, Zhou2021, Guo2021, Martin2022, nie2024single, wang2023deepbasis, zhang2023deep,zhou2022look}, optimization~\cite{Gao2019, Guo2020, zhou2023semi, Henzler2021, luo2024single}, or denoising process~\cite{sartor2023matfusion, vecchio2024controlmat}. As opposed to our goal, these methods target the acquisition of SVBRDFs from input photographs, while our approach focuses on generating new materials from text prompts. 

\paragraph*{Material Generation.} An alternative method to obtaining materials is through material generation. In the industry, the material creation process heavily relies on complex procedural node graphs, which require significant expertise. To simplify the material creation process, various learning-based material generators have been proposed, either using GANs~\cite{Guo2020, zhou2022tilegen} trained on synthetic data or tailoring the training procedure to work exclusively with real materials captured using flash photographs~\cite{zhou2023photomat} for improved realism. An alternative research direction targets procedural material generation, using generic material graphs and user-driven segmentation~\cite{hu2022} or direct generation using transformer architectures~\cite{guerrero2022matformer, hu2023generating}. With the exception of PhotoMat~\cite{zhou2023photomat}, which is trained on a limited dataset, these material generators are trained on synthetic dataset, creating a visual gap to real-world material.

Recently, DMs have emerged as state-of-the-art generative models, leading to their use in multiple material generators. MatFuse and MatGen~\cite{vecchio2024matfuse, vecchio2024controlmat} train an LDMs-based material generator from scratch with multiple encoders conditioned on text, image, or sketch. Closer to our approach, Text2Mat~\cite{he2023text2mat} and ReflectanceFusion~\cite{xue2024reflectancefusion} propose to leverage pretrained diffusion models on large-scale image datasets. These approaches first generate a latent representation of a natural image or center-flashed material image through pretrained DMs or finetuning DMs. At the second stage, they train a separate decoder to extract SVBRDF parameters, as done in the material acquisition task. With this process, these methods benefit from the realistic prior of DMs; however, the reliance on an ``acquisition'' step leads to typical ``baked lighting'' limitations. In contrast, we propose to directly finetune a DM to output material parameters, benefiting from the network's prior without the typical acquisition limitations, and introduce a Reinforcement Learning (RL) approach to improve realism.

\paragraph*{Reinforcement Learning of DMs.}

RL has been widely used in large language models and computer vision tasks. With the success of DMs in text-to-image generation applications, several RL-based algorithms have been proposed to improve the DMs with a goal of finetuning a model to achieve higher reward instead of fitting a target distribution. More specifically, Lee at al.~\shortcite{lee2023aligning} incorporate human feedback to improve text-to-image alignment of DMs. They finetune DMs on fixed dataset sampled from pretrained DMs using reward-weighted regression (RWR) method, where denoising process is reframed as one-step Markov Decision Process (MDP). Built upon RWR, Dpok~\cite{dpok} utilizes new samples from pretrained DMs and include KL-regularization during finetuning process. Both RWR and Dpok are designed from a specific training prompt each training loop and rely on approximate log-likelihood due to single-step MDP. In comparison, denoising diffusion policy optimization (DDPO)~\cite{ddpo} treats the denoising process as a multi-step MDP, and thus is more accurate than RWR and can incorporate many training prompts at once. Xie et al.~\shortcite{xie2024carve3d} have demonstrated that DDPO can be applied to improve 3D consistency of generated multi-view images. To improve realism of a material generator, we build \method{} upon DDPO, with new base model and self-defined realism reward trained using large-scale real dataset.

\section{Preliminaries}
\label{sec:preliminaries}

\begin{figure*}
    \centering
    \includegraphics[width=1\linewidth]{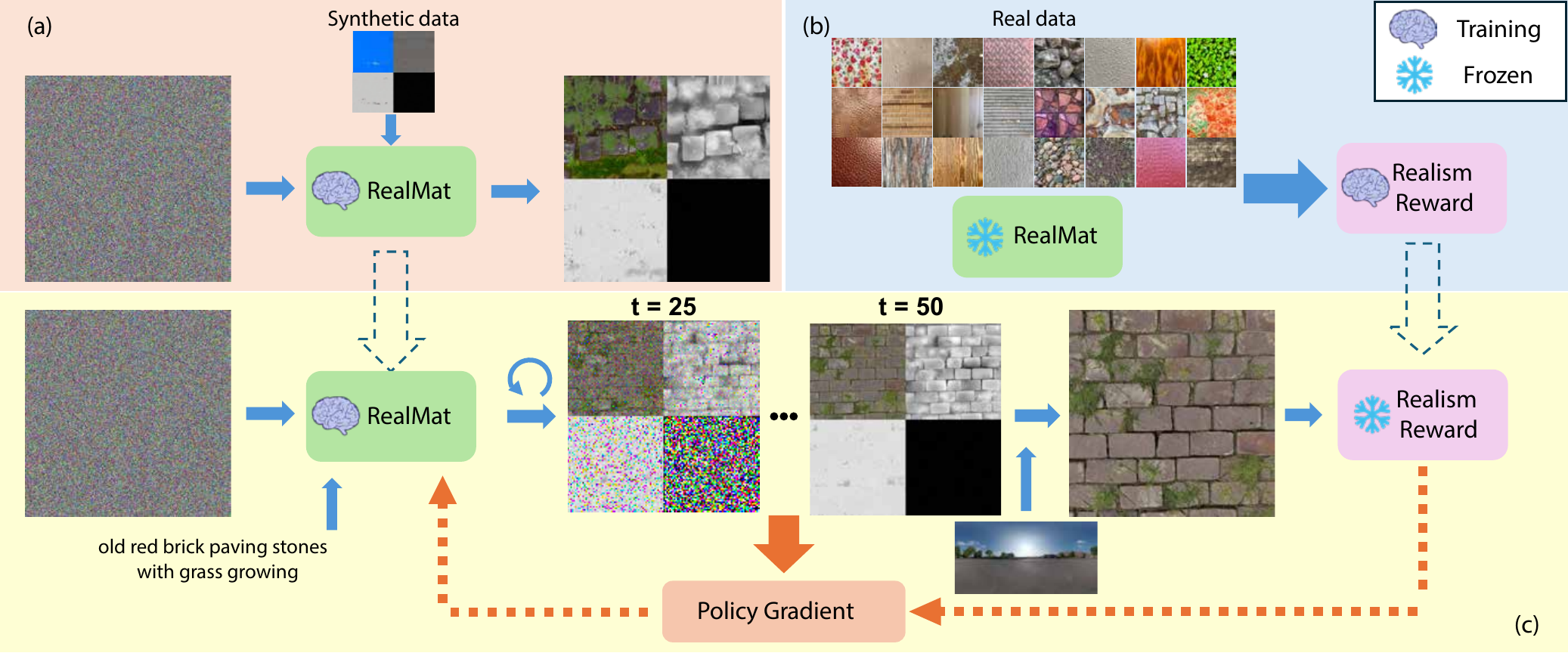}
\caption{ A visual illustration of \method{}. (a) In the first stage, we finetune SDXL for text-to-material generation using synthetic SVBRDF maps arranged in $2 \times 2$ grids;
(b) next, we train a realism reward (score) using a mixture of real photographs and synthetic data;
(c) in a second finetuning stage, we use the reward in a reinforcement learning (RL) strategy
to further push the generated distribution towards more realistic materials.}
\vspace{-0.1in}
\label{fig:realmat}
\end{figure*}

In this section, we briefly introduce two techniques that are central to our approach: image diffusion models~\cite{ho2020denoising} and DDPO~\cite{ddpo}.

\subsection{Diffusion Models}
\label{sec:diffusion_models}

Diffusion models~\cite{ho2020denoising} represent a distribution $p(x_0 | c)$ of images $x_0$ conditioned on $c$ (e.g. a text prompt embedding) by learning to approximately invert a \emph{forward process} $q(x_t | x_0)$ that 
maps images $x_0$ to noisy versions of the images $x_t$, with a noise strength given by a \emph{time step} $t \in [0, T]$.
To invert this mapping, a network $f_\theta(x_t, t, c)$ is trained with the following loss to approximate $x_0$ given a noisy image $x_t$, the time step $t$, and the condition $c$:
\begin{equation} \label{eq:denoising_loss}
    \mathbb{E}_{(x_0, c) \sim p(x_0, c), x_t \sim q(x_t | x_0), t \sim \mathcal{U}(0,T)}\left[\|f_\theta(x_t, t, c) - x_0\|^2\right].
\end{equation}
Images $x_0$ can then be obtained by iteratively sampling small denoising steps. We use a Markovian (i.e. probabilistic) DDPM sampler~\cite{ho2020denoising}:
\begin{equation} \label{eq:denoising_step}
p_\theta(x_{t-1} | x_t, t, c) = \mathcal{N}(a_t x_t + b_t f_\theta(x_t, t, c),\ \sigma_t^2 \mathbf{I}),
\end{equation}
where the factors $a_t$ and $b_t$, as well as the variance $\sigma_t^2$ are chosen according to the denoising schedule of the sampler, and determine a step size that trades off between quality and total number of steps needed. The sequence of images $\tau = (x_T, \dots, x_0)$ forms a denoising trajectory in image space that starts with pure noise $x_T$ and ends with the fully denoised image $x_0$. 

Our diffusion model is based on Stable Diffusion XL (SDXL)~\cite{podell2023sdxl}, a latent diffusion model where images $x$ are represented in a latent space. This model can handle high-resolution $1\text{k} \times 1\text{k}$ generation. We additionally use Zero-SNR~\cite{lin2024common}, a technique to further improve inference quality.

\subsection{Denoising Diffusion Policy Optimization}
\label{sec:ddpo}

%
Recently, it has been shown that Proximal Policy Optimization (PPO)~\cite{schulman2017proximal}, a form of reinforcement learning, can be applied to refine diffusion models in an approach called 
DDPO~\cite{ddpo}. 
DDPO allows us to optimize diffusion models to maximize a potentially non-differentiable reward that can be computed on the image we obtain after multiple denoising steps (Eq.~\ref{eq:denoising_step}). This contrasts with the traditional loss used to train diffusion models, which optimizes only a single denoising step (Eq.\ref{eq:denoising_loss}).

In PPO, a \emph{policy} is trained to perform \emph{actions} that maximize the \emph{accumulated reward} over a sequence of actions. In DDPO, the policy corresponds to the denoising steps $p_\theta(x_{t-1} | x_t, t, c)$ and actions correspond to images $x_{t-1}$ from the denoising trajectory, so that the denoising trajectory $\tau = (x_{T-1}, \dots, x_0)$ is the sequence of actions that results in the denoised image $x_0$. DDPO then aims at maximizing the expected accumulated reward $R(\tau)$ over denoising trajectories $\tau$:
\begin{equation} \label{eq:expected_accumulated_reward}
    J(p_\theta) \coloneq \mathbb{E}_{\tau \sim p_\theta} \left[ R(\tau) \right],
\end{equation}
In DDPO, $R(\tau)$ is computed based on only the final denoised image $x_0$ (i.e., no reward is given for intermediate denoising steps). We will define the exact reward we use in
Section~\ref{sec:realism_refining}.

To optimize this objective, we follow the formulation of Schulman et al.~\cite{schulman2017proximal}, which shows that the gradient of $J(p_\theta)$ w.r.t. the parameters $\theta$ is:
\begin{equation} \label{eq:policy_gradient}
    \nabla_\theta J(p_\theta) = \mathbb{E}_{\tau \sim p_\theta} \left[\sum_{t=T}^1 R(\tau) \nabla_\theta \log p_\theta(x_{t-1} | x_t, t, c) \right].
\end{equation}
The expectation in this equation is approximated using multiple trajectory samples, and the resulting gradient is used to finetune the denoiser parameters $\theta$. Optimization with this gradient can further be stabilized using trust region clipping and importance sampling strategies, as described in DDPO~\cite{ddpo}. Note that the gradient of $R$ is not needed in the computation.

\section{\method{}}
\label{sec:method}

Our goal is to train a diffusion-based SVBRDF generator that produces high-fidelity, diverse, and photo-realistic materials from text prompts.
The main challenge in this task is to preserve priors from existing data sources. \emph{Pre-trained diffusion models} like Stable Diffusion XL \cite{podell2023sdxl} offer great text alignment, diversity, and realism, but are meant for general RGB images and cannot generate material maps. \emph{Synthetic material datasets} \cite{Substance,vecchio2024matsynth} contain ground truth SVBRDF maps, but are limited in size and do not fully bridge the realism gap. \emph{Photographed material datasets} \cite{zhou2023photomat} offer realism, but do not have SVBRDF ground truth and are limited in size.

Our key insight is that the large diffusion models and the combination of real and synthetic data can be used to generate realistic SVBRDFs. 
\xz{To achieve this, we propose a two-stage approach, where the first stage bootstraps SDXL for material generation, and the second stage focuses on realism improvement. Specifically,} We first finetune SDXL~\cite{podell2023sdxl} to generate SVBRDFs using a synthetic material dataset (Section~\ref{sec:synthetic_finetuning}). This model
inherits the text alignment, diversity, and realism from SDXL, but the tuning process moves the distribution towards a more synthetic appearance, creating a realism gap. 
\xz{In the second stage, we additionally leverage a dataset of material images captured under natural (non-flash) lighting to design a realism reward function. This reward function is then utilized to evaluate renderings using sampled materials after multiple denoising steps and to supervise the model finetuning via reinforcement learning with DDPO (Section~\ref{sec:realism_refining}). Our renderer in RealMat is set to be non-differentiable, although a differentiable renderer could also be supported using alternative reinforcement learning frameworks.} We show an overview of our approach in Fig.~\ref{fig:realmat}.




\subsection{Datasets}
\label{sec:datasets}

\paragraph*{Synthetic Data.} We follow previous work~\cite{vecchio2024controlmat} and use $8,615$ material graphs to generate $\sim 126,000$ material variations which we render under randomly selected natural environment maps to create $\sim 800,000$ pairs of renderings and material properties. We augment this data through crops and rotations. This data is used in our first stage of finetuning SDXL.



\paragraph*{Generated Materials.} Following our first stage of finetuning, we can easily sample thousands of materials from text. These materials, however, demonstrate varying degrees of realism and \xz{require an additional labelling process before realism reward training}. \xz{To construct a dataset for labelling}, we randomly sample 2,000 materials, each of which is rendered under 10 random environment lighting, for a total of 20,000 renderings in our Generated Materials dataset. \xz{Our realism function is trained on environment-lit materials instead of flash-lit materials to be compatible with large real data.} 

\paragraph*{Real Photographs.} We filter 800,000 material photographs from an internal image database through a nearest neighbor CLIP-based search using FAISS~\cite{johnson2019billion}. Note that although the database contains high-quality images in general, not all material samples are guaranteed to be realistic. Therefore, \xz{similar to generated material data}, a classifier is still necessary to filter out realistic data from the Real Photographs dataset. 


\paragraph*{Data labeling.} We filter the Generated Materials and Real Photographs datasets for realism. Because manually annotating both datasets entirely is impractical, we propose to train classifiers separately on each dataset to estimate the realism labels. Initially, each classifier is trained on a small subset of 2000 images manually annotated as realistic or unrealistic. To prevent overfitting to these small training sets, we employ simple linear classifiers that take CLIP embeddings of the images as input and produce probabilities indicating realism. After training, each classifier predicts realism scores for the remaining unlabeled images in its respective dataset. We then apply empirically determined thresholds (0.2/0.4 for the Real Photographs/Generated Materials) to generate binary realism labels. The labels are used to train our realism reward function in Stage 2 (Section~\ref{sec:realism_refining}).


\subsection{Stage 1: Finetuning SDXL}
\label{sec:synthetic_finetuning}

We choose SDXL, trained on the large natural image dataset Laion-5B~\cite{schuhmann2022laion}, as it offers good realism, diversity, and text alignment, and is able to generate $1\mathrm{k} \times 1\mathrm{k}$ images; it also empirically responds well to the RL stage below. 

Inspired by the 3D generation method~\cite{li2023instant3d}, in which four multi-view images of an object are generated as a grid, we represent SVBRDFs as a $2 \times 2$ grid: albedo, height, roughness and metallic. We then finetune SDXL using these SVBRDF grids. In practice, this lets us trade spatial resolution for the capacity to generate more channels. \xz{Due to the global attention mechanism in SDXL, each feature patch can get attention from both nearby and distant feature patches, enforcing pixel-level consistency across the generated $2 \times2$ material layout (as shown in Fig.~\ref{fig:base_SDXL}).}

Although our base model inherits much of the diversity and realism of SDXL, the finetuning process still moves the distribution towards a more synthetic appearance. We demonstrate this behavior in Fig.~\ref{fig:base_SDXL}, with synthetic (top row) and realistic examples (bottom row) generated after this finetuning stage.

\subsection{Stage 2: Finetuning for Realism}
\label{sec:realism_refining}

In Stage 2, we refine our material generator to close the realism gap created in Stage 1. The unknown setting the photos were taken in (lighting conditions, camera parameters, etc.) makes any attempt to use a differentiable renderer to supervise renders of the generated SVBRDFs difficult. \xz{ Although we could leverage differentiable and limited multi-step finetune strategy to train our models, this could trigger significant computational overhead, and using a reduced number of inference steps may further degrade generation quality. Instead, we design a realism reward function, which is set as non-differentiable, and we propose to use an RL strategy to finetune our material generator with the goal of maximizing the realism reward.} 

\paragraph*{Realism reward function.}
The reward function has the same architecture as our realism classifiers for datasets filtering, without the sigmoid activation after the last layer since we target a reward score rather than a classification. Here again, we use the CLIP~\cite{radford2021learning} features as input as they have been shown to perform well on materials~\cite{Yan:2023:PSDR-Room} and have been used to define a general image aesthetic score~\cite{ddpo}. In summary, our score function $r_\phi(I)$ is a linear layer, mapping CLIP features to a realism score.
We use an MSE loss to train the reward function, with an additional Total Variation (TV) \cite{johnson2016perceptual} regularization. \xz{In TV loss, we obtain the transformed input images $I^{\prime}$ by applying rotation/scaling/translation to the original input images $I$. Then we minimize the $\mathcal{L}_2$ distance of the CLIP features between $I$ and $I^{\prime}$, encouraging similar scores for similar inputs}. Our complete loss $\mathcal{L}_r$ for the reward function training and TV loss are formulated as:


\begin{equation}
\mathcal{L}_r = \lambda_1 \Vert r_\phi(I) - l_{gt} \Vert_{2} + \lambda_2 \mathcal{L}_{tv} \big( \text{CLIP}(I)\big) .
\label{eq:tv_loss}
\end{equation}

\begin{equation}
\mathcal{L}_{tv} = \Vert \text{CLIP}(I) - \text{CLIP}(I^{\prime}) \Vert_{2}.
\label{eq:tv_loss_1}
\end{equation}

\noindent where $l_{gt}$ represents the thresholded binary realism labels \xz{(0 and 1)}, $\mathcal{L}_{tv}$ is the TV loss term, and
$\lambda_1$ and $\lambda_2$ weight the MSE and TV loss terms. 
The accumulated reward $R$ in Eq.~\ref{eq:policy_gradient} is then defined as:
\begin{equation}
    R(\tau) \coloneq r_\phi\Big(g\big(d(x_0), L\big)\Big) \text{ with } L \sim \mathcal{U}(\mathbf{L}),
\end{equation}
where $x_0$ is the final denoised latent image, $d$ is the VAE decoder, and $g(X, L)$ renders material maps grid $X$. The rendering uses a lighting environment $L$, sampled uniformly from a set $\mathbf{L}$ of 200 natural lighting environments. Note that rewards for intermediate diffusion steps are not needed. 

\paragraph*{Training prompts.} To ensure that RL training improves realism where it is most needed, we design a set of ``training prompts'' which are sampled during the training. We first take $1,000$ text prompts from the image descriptions, covering 16 material categories (brick, ceramic, fabric, flower, granite, grass, ground, leather, leaves, marble, metal, paper, pebble, rock, stone and wood). For each text prompt, we sample ten materials using our finetuned material SDXL and evaluate their realism score with our realism reward function. We then select $6$ or $7$ prompts with the lowest realism within each category, for a total of $100$ prompts. This ensures improved realism across a sufficiently diverse range of appearances, especially for concepts that typically lacked realism before RL. We include the list of prompts in the supplemental materials and evaluate the effect of changing the number of training prompts in Section~\ref{sec:ablation_study}.

\section{Implementation}
\label{sec:impl}


\begin{figure}
    \centering
    \includegraphics[width=1\linewidth]{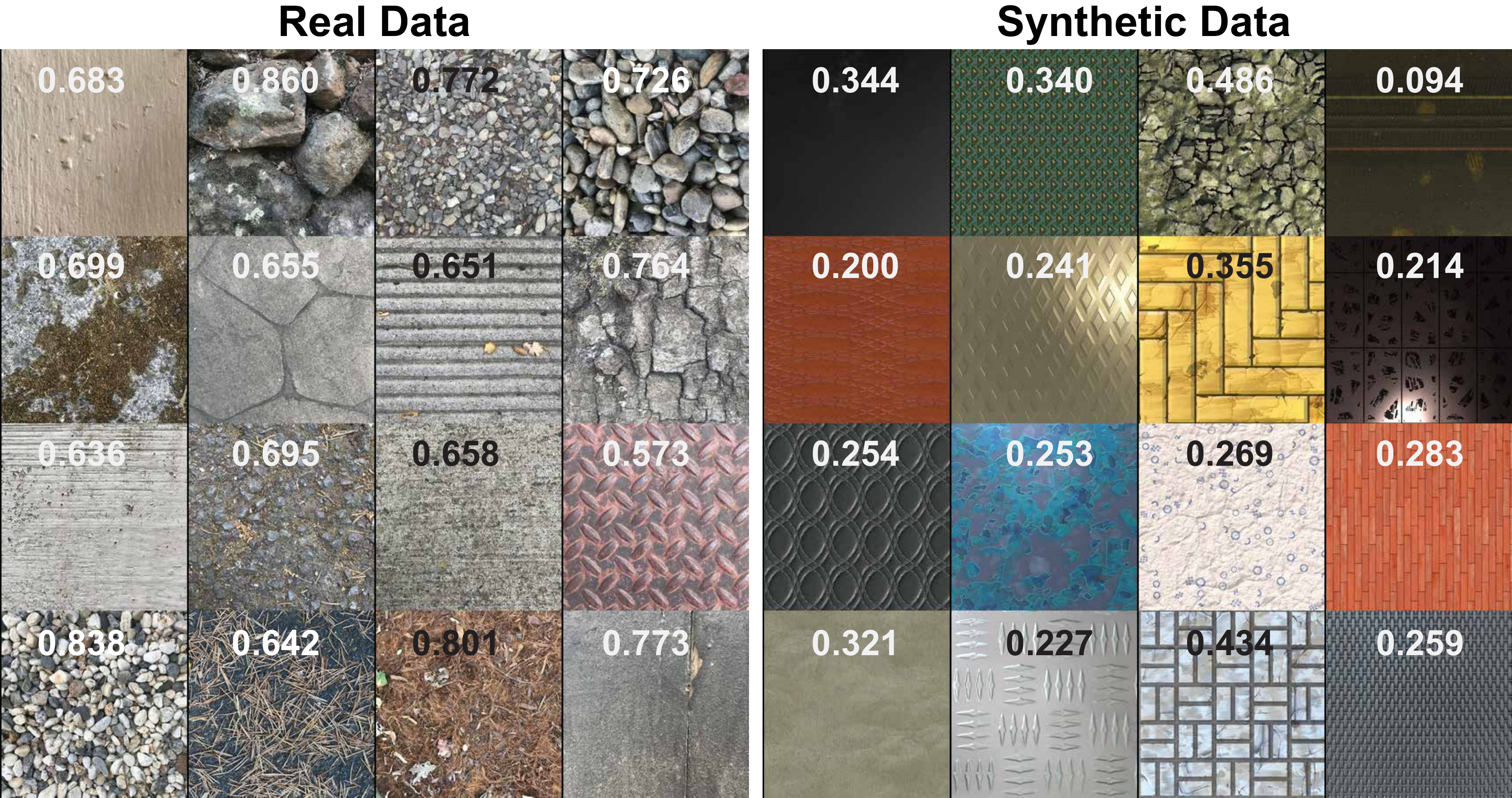}
\caption{The estimated normalized realism scores from our realism reward function. Left shows the scores of real materials and right shows the scores of rendered synthetic materials, demonstrating the effectiveness of the realism reward function.}
\label{fig:score_realsyn}
\end{figure}

\paragraph*{SDXL finetuning.} We finetune with the $1024 \times 1024$ material grids from the Synthetic Data. We create corresponding text prompts by combining material category names (e.g.``wood'' or ``metal'') with the material names. We finetune SDXL using AdamW \cite{loshchilov2017decoupled} with a learning rate of $\num{2e-6}$ and batch size of 120. We set terminal signal-noise ratio (SNR) to zero~\cite{zeroSNR}, which helps to generate a high-contrast range (e.g. colorful albedo and solid black metallicity in the same grid). We finetune SDXL for 400k iterations ($\sim 7$ days) on 24 80GB A100 GPUs.

\begin{figure}
    \centering
    \includegraphics[width=1\linewidth]{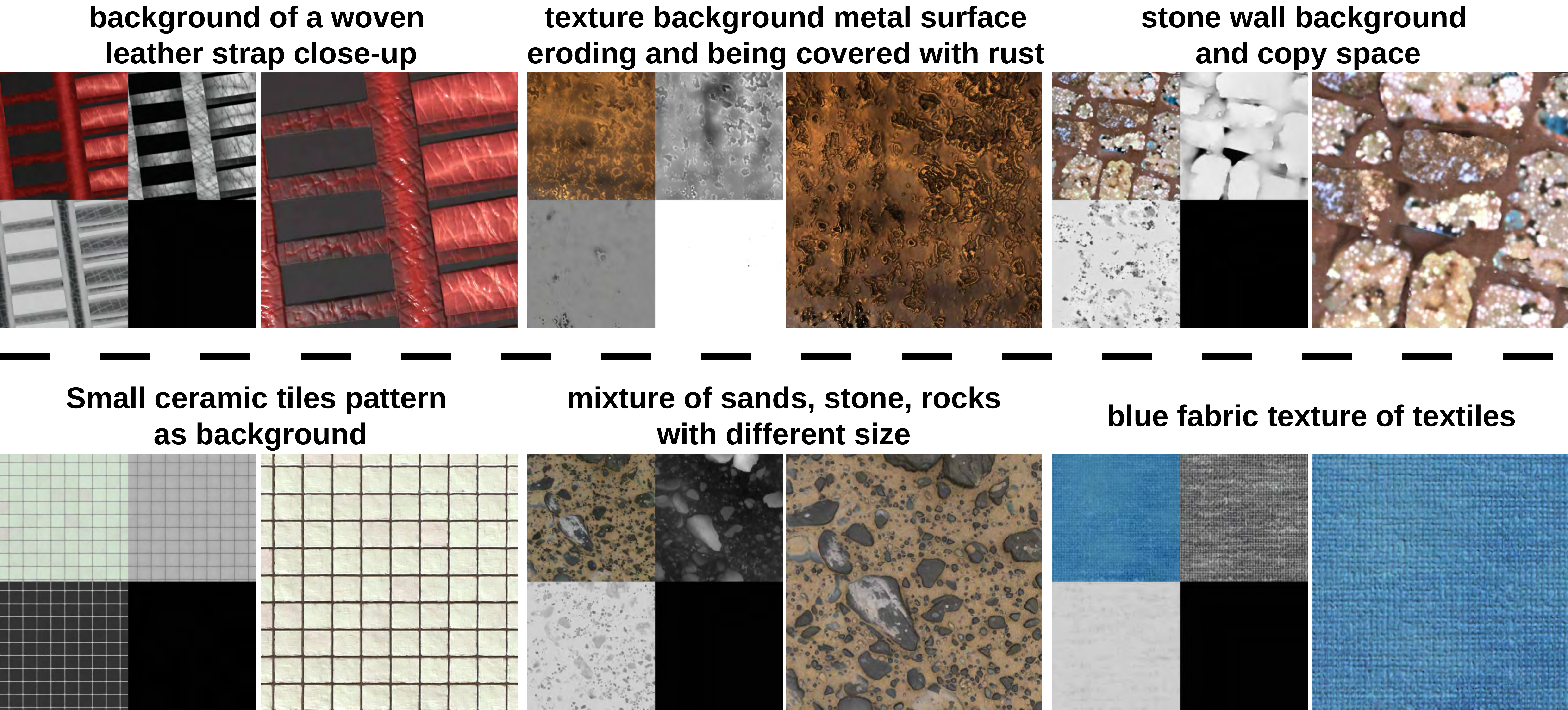}
\caption{The sampled material maps and renderings of \method{} at the first stage of fine-tuning with the synthetic dataset. Here, we partition examples as synthetic (top row) and realistic (bottom row) to motivate our reinforcement learning realism fine-tuning stage (Sec.~\ref{sec:realism_refining}). In all cases, the materials are high-quality, show consistent features across different material maps, and preserve good text alignment.}
\label{fig:base_SDXL}
\end{figure}

\paragraph*{Training realism reward function.} 
We train our realism classifiers and realism reward function with a learning rate of $\num{1e-3}$ and batch size of 64. The weights of MSE term $\lambda_1$ and TV term $\lambda_2$ are set as $1$ and $100$ respectively. The training takes around 6 hours on a single 40GB A100 GPU.


\begin{figure*}
    \centering
    \includegraphics[width=1\linewidth]{figs/Progressive_c.pdf}
\caption{This figure shows materials resulting from increasing levels of RL fine-tuning with realism reward, starting from no fine-tuning (left) to fine-tuning for 110 epochs (right). RL fine-tuning progressively improves synthetic materials (top three rows), while remaining consistent for realistic materials (last row). This demonstrates the effectiveness and robustness of the second stage of \method{}.}
\label{fig:progressive_DDPO}
\end{figure*}



\paragraph*{RL training.} We use 32 80GB A100 GPUs. We use 50 denoising steps. We use a batch size of 128 (4 per GPU) and gradient accumulation over two minibatches.
Similar to DDPO, instead of finetuning the entire base model, we freeze it and only finetune a Low Rank Adaptation (LoRA) \cite{hu2021lora} with rank 4 to reduce memory and computation cost. We finetune the model with RL for 110 epochs ($\sim 18$ hours) with a fixed learning rate of $\num{3e-4}$.

\paragraph*{Rendering.} We use Mitsuba 3 \cite{Mitsuba3} to render materials on fully displaced meshes under 200 randomly selected realistic HDR environment maps from Poly Haven \cite{envHDR}. The environments are normalized to preserve the same irradiance on a flat plane per RGB channel, to prevent overall intensity and color differences.

\section{Results and Discussion}

\begin{figure}
    \centering
    \includegraphics[width=1\linewidth]{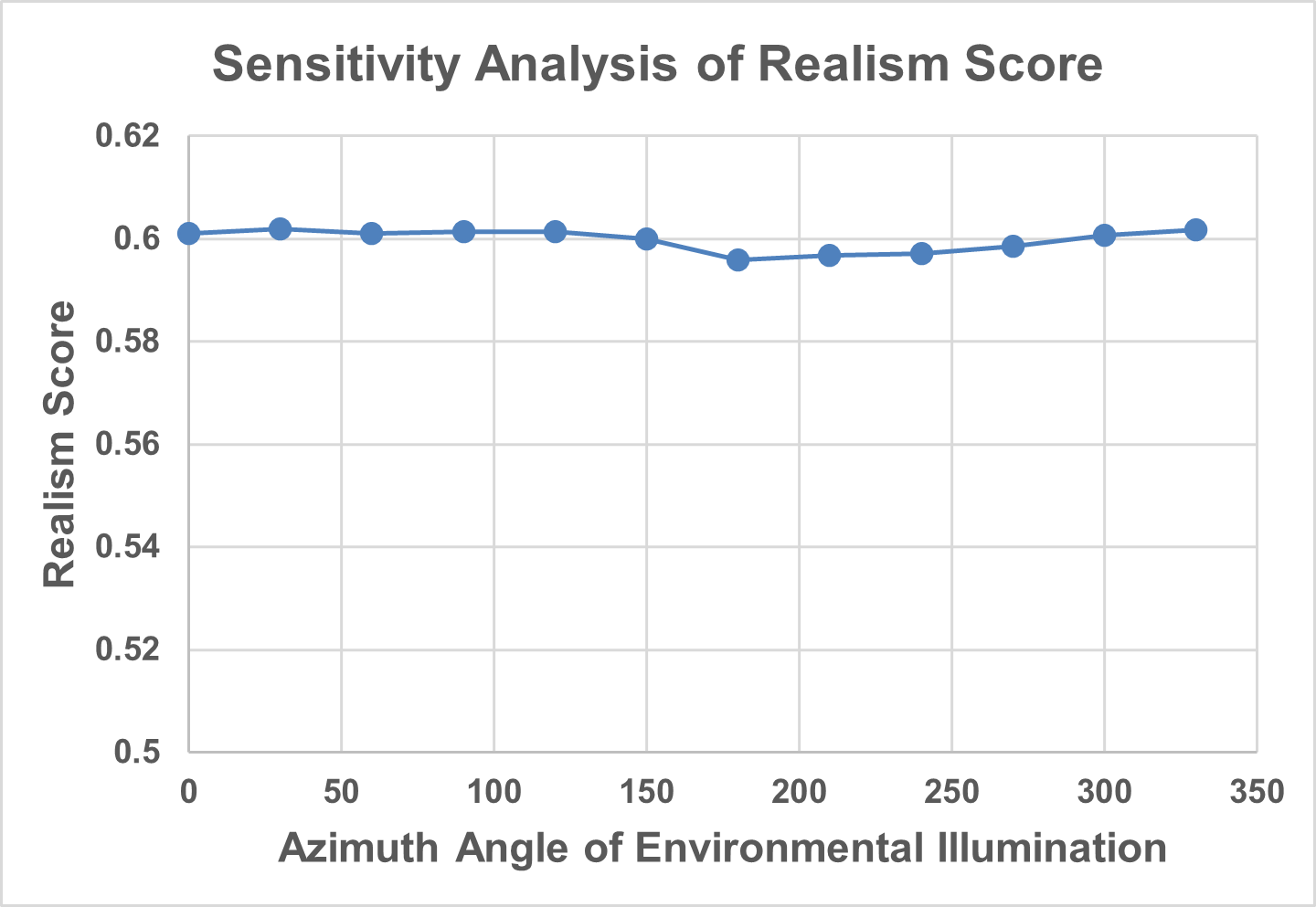}
\caption{Plot of realism score under illumination with different azimuth angles. We render 100 randomly selected materials under 20 different environment maps, and average out the score per rotation angle across all renderings.}
\vspace{-0.2in}
\label{fig:rebuttal_plot}
\end{figure}

In this section, we first evaluate our realism reward function numerically and visually. Next, we show sampled results of \method{} from both stages, and perform a RL finetuning user study to evaluate the effect of the second finetuning stage. In addition, we ablate the effect of training prompt number on the realism RL finetuning process and the effect of TV regularization in the reward function. Finally, we compare \method{} against previous methods qualitatively and through a second user study, demonstrating that our method can generate realistic and diverse materials.

\subsection{Realism Reward Function}

We first evaluate the realism reward function on both real and synthetic test sets. We find that on average, the reward function estimates a realism score of 0.73 for a real test set (photographs taken by us) and 0.516 for a synthetic test set from Vecchio et al.~\cite{vecchio2024matsynth}. After normalization to the range [0,1] based on max/min values (0.878/0.343), the average scores for the real and synthetic sets are 0.723 and 0.324, respectively. \xz{We further analyze the sensitivity of the score function under different environmental illuminations. Specifically, we render 100 randomly selected materials under 20 different environment maps with varying azimuth angles. We obtain an average realism score of 0.60, with standard deviations of 0.0291 across different environment maps and 0.0156 across rotation angles, respectively. Furthermore, we plot the relationship between the realism score and the azimuth angle, as shown in Fig.~\ref{fig:rebuttal_plot}.} This shows that our reward function is insensitive to illuminations, and successfully acts as a material realism estimator. In addition, in Fig.~\ref{fig:score_realsyn} we show visual examples from real and synthetic test sets with the estimated scores.

\subsection{Results of \method{}}

\paragraph*{Finetuning with Synthetic Data} 
We show materials and corresponding renderings sampled after the first stage finetuning in Fig.~\ref{fig:base_SDXL}. Our model inherits the diversity and details of SDXL, and can often generate high-quality $512 \times 512$ materials with strong text alignment and coherent material channels. Still, this stage creates some realism gap, moving the distribution toward a more synthetic appearance, which motivates the second finetuning stage.


\begin{figure}
    \centering
    \includegraphics[width=1\linewidth]{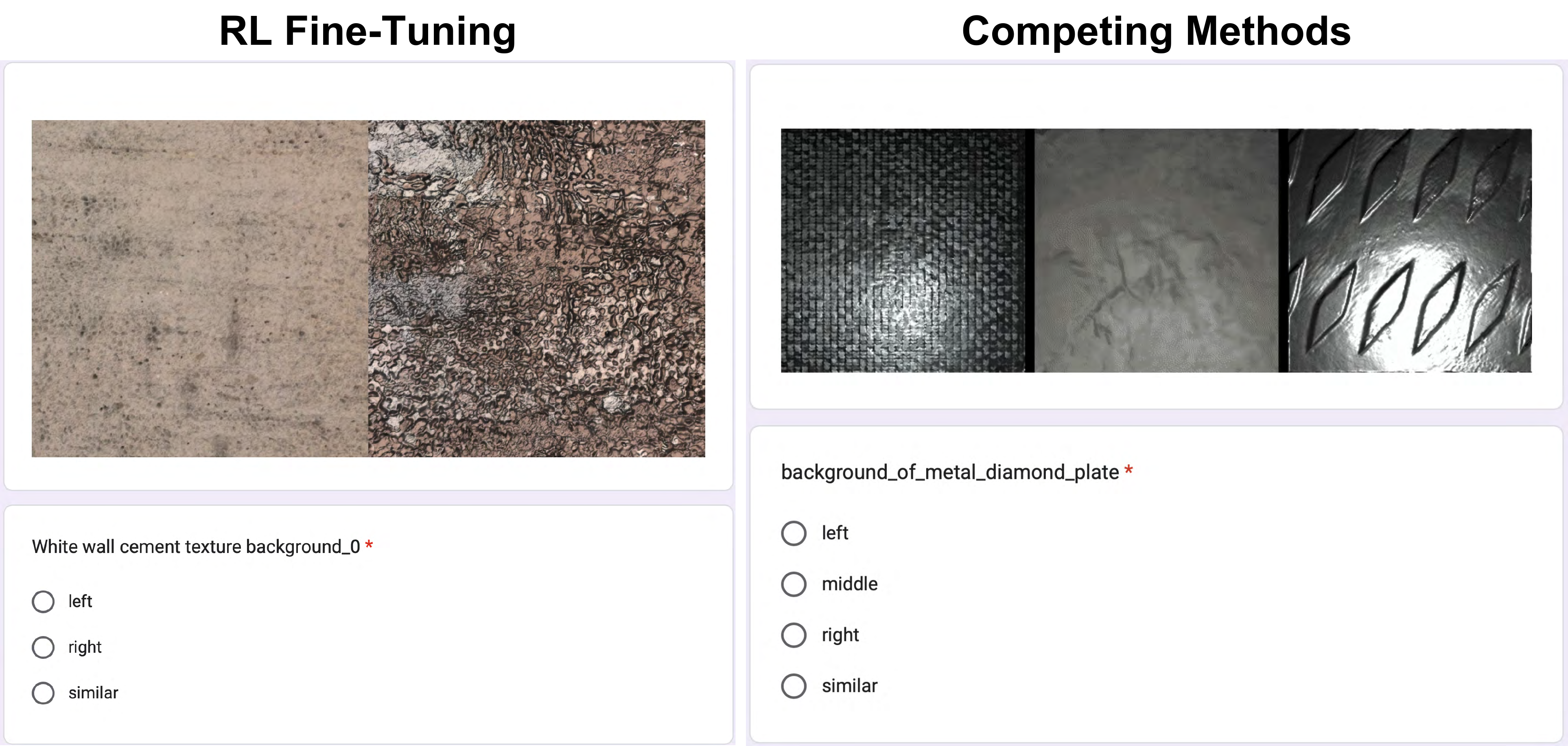}
\caption{The user interfaces of our user studies. For the RL fine-tuning study, text prompts and material pairs are shown to participants who need to select the most realistic option by selecting left, right, or similar. Similarly for the competing methods user study, participants select the most realistic material from three randomized options generated by MatFuse~\cite{vecchio2024matfuse}, ReflectanceFusion~\cite{xue2024reflectancefusion}, and \method{} by selecting left, middle, right, or similar.}
\label{fig:user_study}
\end{figure}

\begin{figure}
    \centering
    \includegraphics[width=1\linewidth]{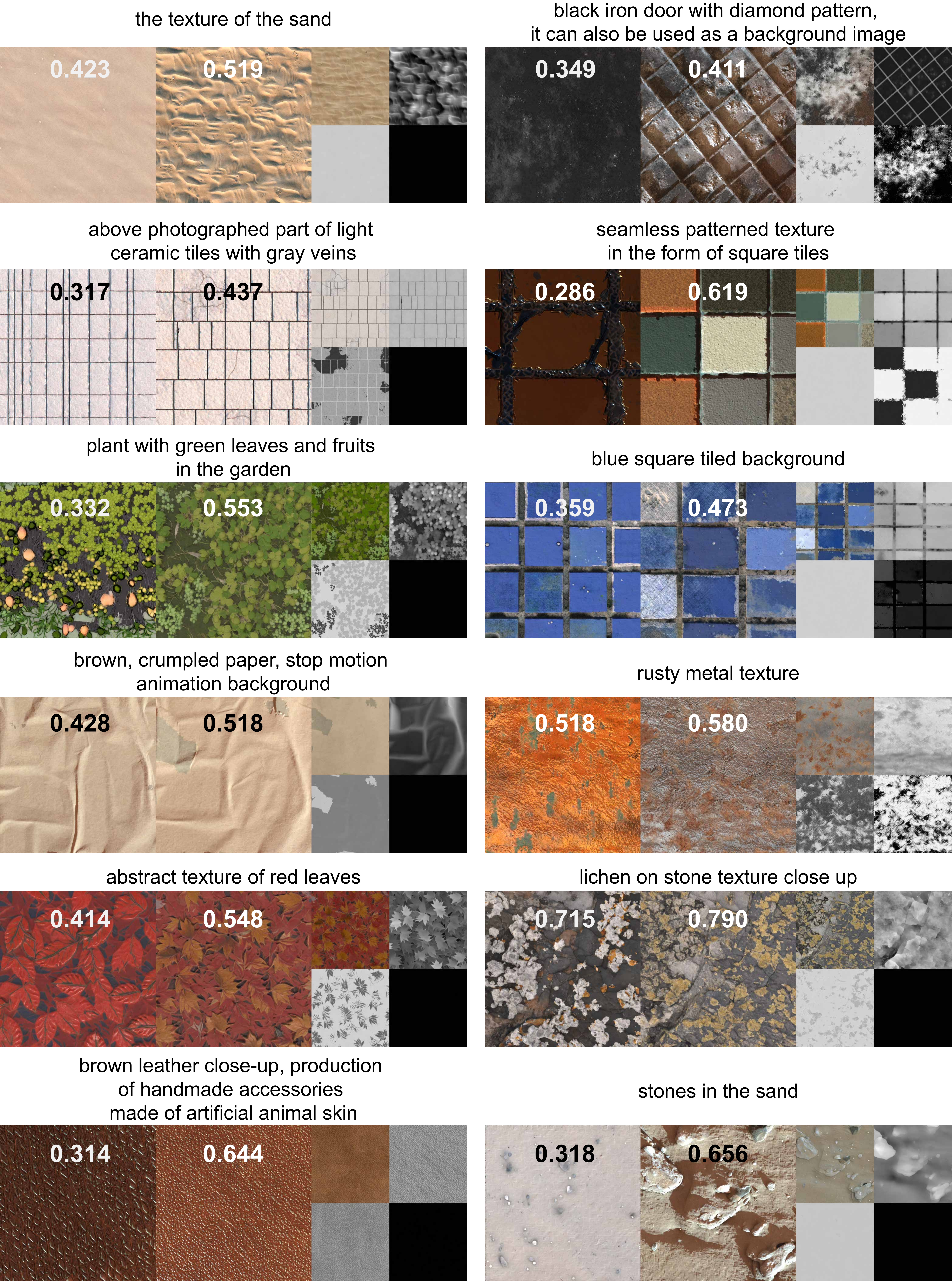}
\caption{We show a side-by-side comparison of renderings before and after RL finetuning, together with the corresponding material maps, given various text prompts. In addition, we show the realism rewards. This demonstrates the impact of our training strategy both numerically and visually. }
\vspace{-0.2in}
\label{fig:step0_110}
\end{figure}

\paragraph*{RL Finetuning for Realism.} 
We show the progressive improvement of \method{} during the RL finetuning with our realism reward in Fig.~\ref{fig:progressive_DDPO}. The leftmost image in each row is a rendering of a material sampled from the model after the first stage of \method{}, before RL finetuning, and the rightmost image after the RL finetuning is complete. We keep the prompt and seed fixed in each row. We can see that the realism and visual appearance of synthetic materials (first three rows) gradually improve from left to right as the RL finetuning progresses. The last row, which is already realistic after the first finetuning stage, remains realistic throughout the RL finetuning.
We also show a side-by-side comparison of renderings in the initial and final RL finetuning step, along with the corresponding realism score in Fig.~\ref{fig:step0_110}. The renderings of materials sampled after RL finetuning display more realistic features such as natural-looking leaves, wavy sand, and cracks in tiles. 


\paragraph*{RL finetuning user study.} \xz{Note that RealMat is conditioned on the text
prompt rather than an image, therefore we cannot perform quantitative realism evaluation using measured real data.} To evaluate the impact of our realism RL finetuning, we conduct a user study asking participants to assess the realism of paired materials. In our user study interface (see left in Fig.~\ref{fig:user_study}), given a material pair and its text prompt, the user is asked to select which material looks more realistic: "left", "right" or "similar". These paired materials are rendered under the same lighting, with one material sampled from the model at the initial step and the other at the final RL fine tuning step, with the same random seed. The materials of each pair are randomly ordered. We have in total 21 sub-surveys, each of which contains approximately 20 material pairs. Participants can take more than one sub-surveys if they desire.

In total, we collect 32 sub-survey responses, corresponding to 641 individual feedback points. On average, $46.8\%$ of responses prefer the materials after realism finetuning, $28.9\%$ of responses hold the opposite opinion, and $24.3\%$ find the results "similar". Among the 421 different material pairs, $43.5\%$ of pairs are marked as improved after finetuning, compared with $23.0\%$ of pairs rated as degraded. The remaining $33.5\%$ of pairs are considered as "similar" or "tie". Note that all the material pairs in the user study are randomly sampled without any data curation. This confirms that RL finetuning improves the realism and visual quality of sampled materials. 
We further compute the 95$\%$ confidence intervals for preference: after finetuning $[42.9\%, 50.7\%]$; before finetuning $[25.4\%, 32.4\%]$ and no preference $[21.0\%, 27.6\%].$


\subsection{Ablation Study}
\label{sec:ablation_study}
\begin{figure}
    \centering
    \includegraphics[width=1\linewidth]{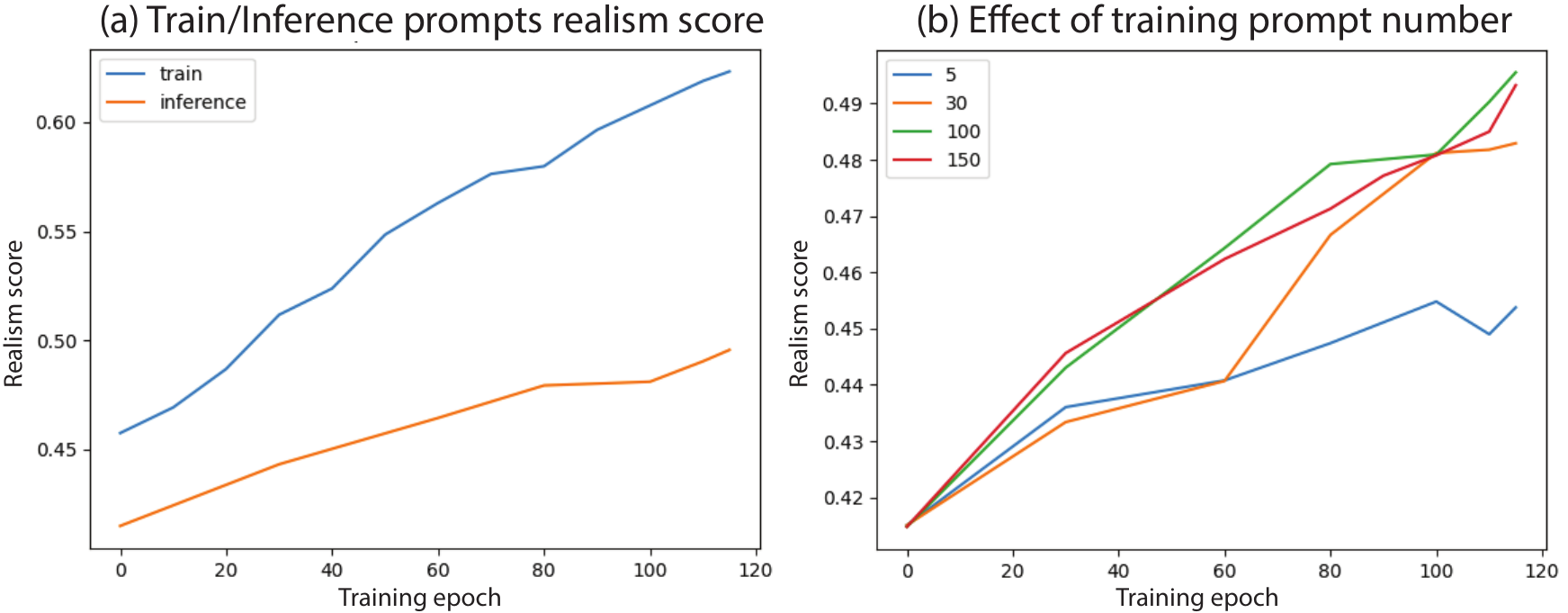}

\caption{Plots of realism score during training: (a) Normalized realism score plot for training and inference text prompts during the RL fine-tuning stage. (b) The effect of training prompt number on the reinforcement learning process. We show the realism score of models trained under 5, 30, 100 and 150 training prompts. Based on the plot, we choose $N=100$ as final optimal training setting.}
\label{fig:plots}
\end{figure}

\paragraph*{Effect of RL} We evaluate how the realism scores of generated materials evolve using both train and test prompts as our RL finetuning progresses in Fig.~\ref{fig:plots} (a). We can see that while materials generated using prompts from the training set benefit most from the RL finetuning, the benefits generalize well to unseen prompts.

\paragraph*{Effect of training prompt set size.} We analyze the effect of training prompt number during the reinforcement learning finetuning stage. Both DDPO~\cite{ddpo} and Carve3D~\cite{xie2024carve3d} claim that finetuning on a limited set of prompts can generalize well to test prompts, which is consistent with our observation. However, in \method{}, we need to carefully distribute the training prompts across all material categories. In Fig.~\ref{fig:plots} (b), we show the effect of training prompt number on realism score. As shown in the figure, even trained on a prompt number as small as 5, the realism still improves as training progresses but appears to be worse than prompt numbers of 30, 100 and 150. Therefore, we choose a training prompt size of 100 in the final setting of \method{} as it achieves the highest realism score based on our analysis.

\begin{figure}
    \centering
    \includegraphics[width=1\linewidth]{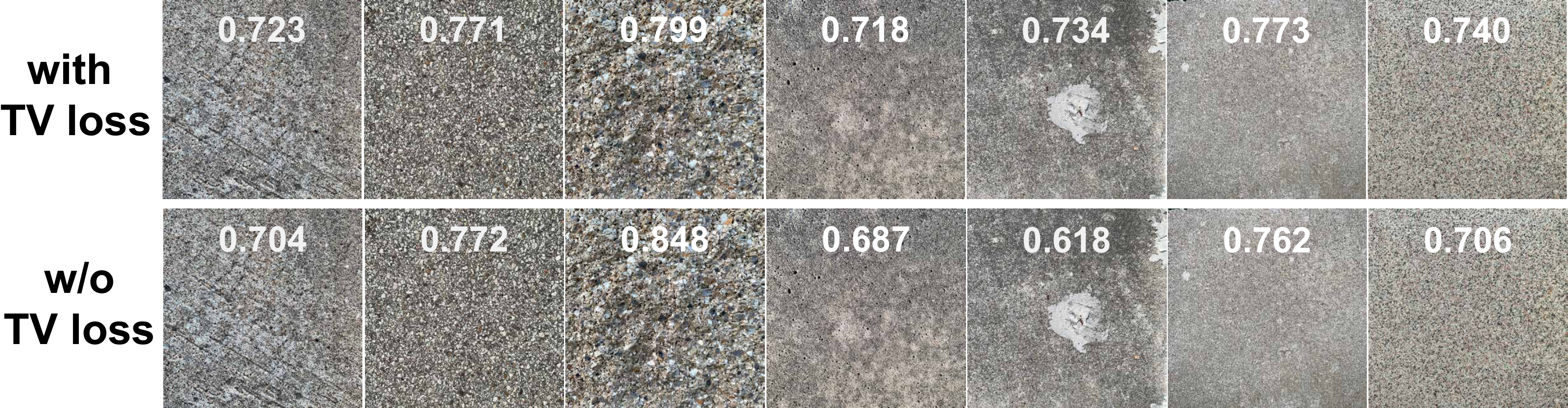}
\caption{We demonstrate the effect of TV regularization when training realism reward function. As demonstrated, using realism reward trained without TV loss, stones with similar appearance yield normalized scores with large variance.}
\label{fig:tv}
\end{figure}

\paragraph*{Effect of TV loss.} TV loss term is added to Eq.~\ref{eq:tv_loss} to regularize the training of the reward realism function, with the goal of ensuring that similar inputs yield similar realism scores. In Fig.~\ref{fig:tv}, we show a set of real stones with similar visual appearance, and the estimated normalized score by realism function with and without TV regularization. These similar stones are expected to output similar scores based on human perceptual assessment. However, without TV regularization (the bottom row in Fig.~\ref{fig:tv}), the maximum and minimum scores are 0.848 and 0.618 respectively, with a standard deviation of 0.074. In comparison, the standard deviation with TV regularization is reduced to 0.030. We conclude that the TV term is helpful to regularize our realism reward function. 

\subsection{Comparison with Previous Methods}

We compare \method{} against ReflectanceFusion~\cite{xue2024reflectancefusion},  MatFuse~\cite{vecchio2024matfuse}, and PhotoMat~\cite{zhou2023photomat}. We used the official code from MatFuse and PhotoMat, and asked the authors of ReflectanceFusion to sample their model with our text prompts.

\begin{figure*}
    \centering
    \includegraphics[width=1\linewidth]{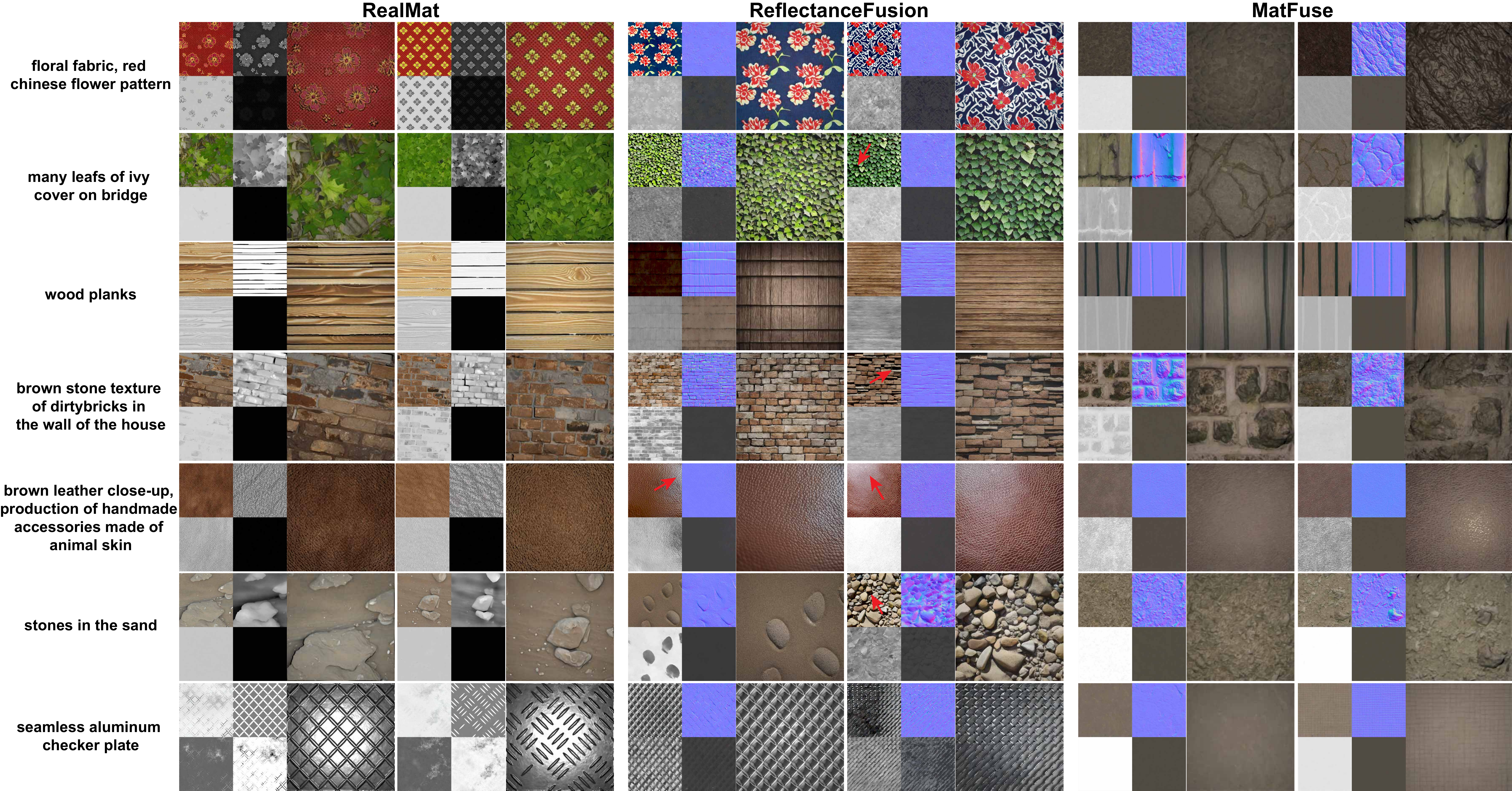}
\caption{Visual comparison between \method{}, ReflectanceFusion and MatFuse. The maps from ReflectanceFusion have light baking artifacts (marked by red arrows), produced by a base (non-finetuned) image model, which are difficult to undo by further processing. MatFuse produces less plausible and diverse materials, due to exclusively synthetic training data. In comparison, RealMat generates diverse, realistic and artifact-free material maps.}
\label{fig:comparison2}
\vspace{-0.2in}
\end{figure*}

\paragraph*{General comparison.} We show materials generated from text prompts rendered under a single point light to compare MatFuse, ReflectanceFusion, and \method{} in Fig.~\ref{fig:comparison2}. As shown, given only a text prompt, MatFuse cannot generate diverse materials. We believe this is largely due to training the model from scratch on synthetic materials, lacking the important priors learned from images that ReflectanceFusion and our approach benefit from. ReflectanceFusion, on the other hand, generates materials that closely follow the input prompt and are overall realistic, but suffer from light-baking artifacts, since the method is based on a base (non-finetuned) SD model that aims to generate RGB images with highlights and shadows, which are non-trivial to undo by the second stage of their approach. In particular, light-baking can be observed in the albedo component (all renderings use a single, centered point light) of the ReflectanceFusion results in Fig.~\ref{fig:comparison2}. In contrast, our approach generates materials with higher photorealism and fewer artifacts, while preserving good prompt alignment and diversity.

\paragraph*{Competing methods user study.} To further validate the perceptual quality of our results, we conduct a user study with competing methods (MatFuse,  ReflectanceFusion, and \method{}) asking participants to assess the realism and prompt alignment of generated materials. We use a similar interface to the RL finetuning user study (right of Fig.~\ref{fig:user_study}), in which participants are asked to choose the most realistic material that also reasonably follows the accompanying text prompt by selecting "left", "middle", "right", or "similar". The randomized materials are rendered with the same animated lighting: a single point light moving along a circle parallel to the material plane. We use GIF animations with 60 frames displayed in a loop to better assess the reflectance behavior of the material as it is illuminated from different directions, and make any baking artifacts clear.

For this user study, we assemble a total of 110 sub-surveys, each containing 10 unique material comparisons. Similarly to the RL finetuning study, participants can take more than one sub-survey. In total, we collect 44 sub-survey responses, corresponding to 440 individual comparisons without any data curation. Here, we show the average followed by the 95\% confidence interval range. 53\% [48.3\%-57.6\%] of responses prefer the materials generated by \method{}, while 40\% [35.6\%-44.8\%] opt for results from ReflectanceFusion. Only 3\% [1.2\%-4.3\%] of responses choose MatFuse, and 4\% [2.2\%-5.9\%] find the materials "similar". Among the 373 unique material prompts, \method{} is considered best for 52\%, ReflectanceFusion comes second with 40\%, MatFuse is chosen for 3\%, and 5\% are ties ("similar"). The study shows that our approach generates more realistic materials than competing methods. It also highlights the impact of leveraging image priors, strategy shared by \method{} and ReflectanceFusion, which combined constitute the overwhelming majority of realistic materials according to the participants.

\begin{figure}
    \centering
    \includegraphics[width=1\linewidth]{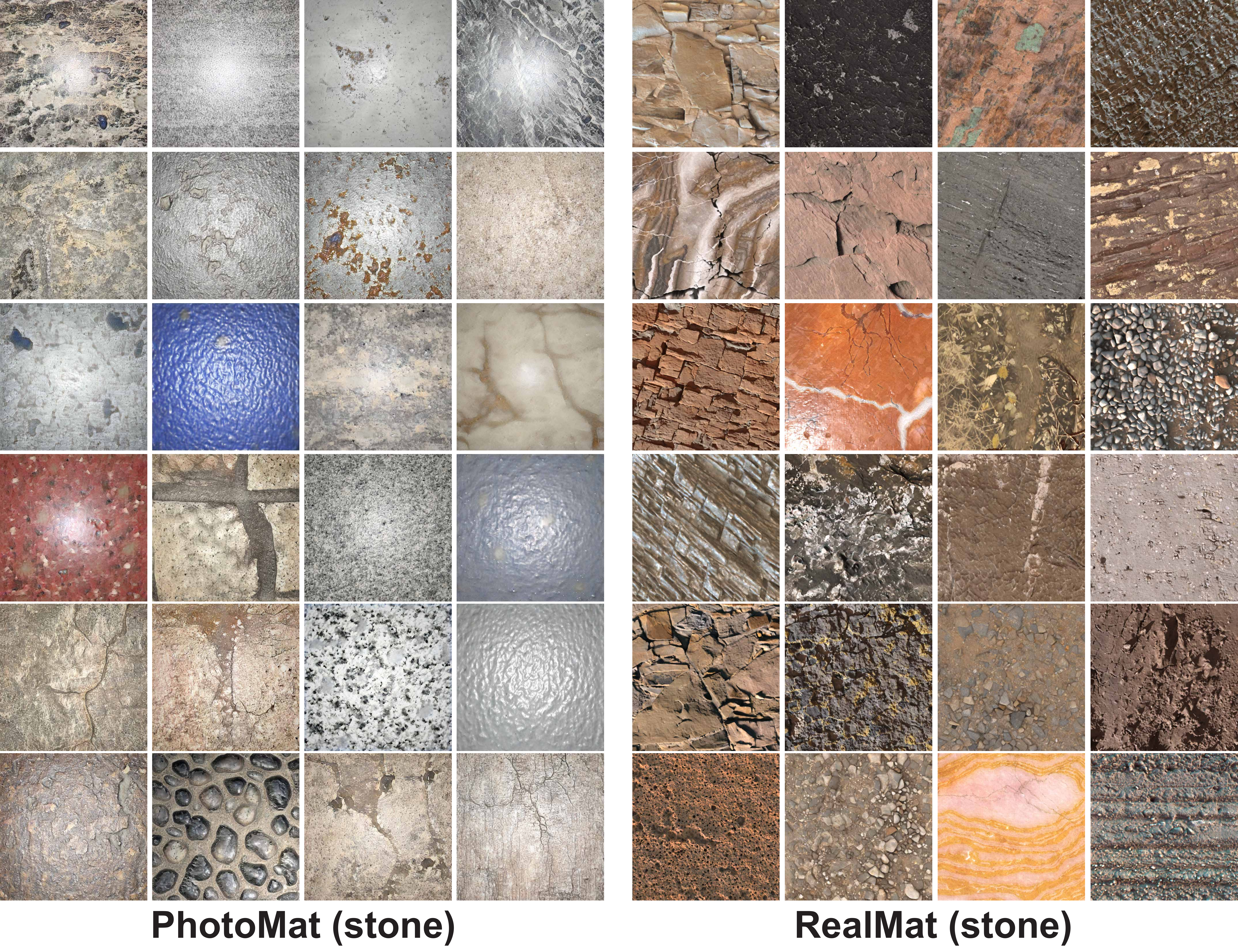}
\caption{Comparison of sampled results of PhotoMat and \method{} (we only focus on one specific category, ``stone''). Although the PhotoMat results are realistic, the sampled materials show limited diversity due to a small-scale real dataset. In comparison, \method{} is trained with realistic priors and demonstrates both strong realism and diversity.}
\label{fig:comparison_photo}
\vspace{-0.2in}
\end{figure}

\paragraph*{PhotoMat comparison.} We compare \method{} with PhotoMat, a material generator that is exclusively trained on real flash photos. Results are shown in Fig.~\ref{fig:comparison_photo}. Note that PhotoMat is only designed for class-conditioned generation, without text control. Due to the limited scale of the PhotoMat training dataset, the sampled materials are fairly realistic but tend to lack diversity (we show sampled ``stone'' of PhotoMat in Fig.~\ref{fig:comparison_photo}). In comparison, \method{} leverages existing realistic priors and produces realistic stones with greater diversity.

\section{Limitations and Discussion}

We observe that detailed text control for material generation remains challenging for precise descriptions (e.g. separate control of the colors of each tile). We believe this may be difficult due to the lack of high-quality text prompts in material datasets. A second limitation of our proposed approach is spatial resolution, which is traded off for additional channels. While this could be mitigated using newer, high-resolution models, it makes the generalization of the grid approach beyond 4 maps more challenging. 

\xz{In principle, our full reward function could be made differentiable by combining a differentiable Mitsuba renderer with a realism score function. This would allow several alternative training frameworks, such as AlignProp~\cite{prabhudesai2023aligning}, DRaFT~\cite{clark2023directly}, ReFL~\cite{xu2023imagereward}, and DRTune~\cite{wu2024deep}, to supervise the final outputs of diffusion models using differentiable reward functions. Note that these methods differ in the sampling of the denoising steps and the gradient truncation. Therefore, exploring the feasibility of these strategies with a differentiable renderer and reward function would be an interesting direction for future work.}


\section{Conclusion}

We propose \method{}, a realistic material generator trained to preserve the prior knowledge in large text-to-image diffusion models and enhance generated material realism. We first adapt the SDXL model to generate materials using finetuning on $2 \times 2$ tiled SVBRDF grids. We further enhance the realism of generated materials by utilizing reinforcement learning (RL) with a realism reward function designed on a large-scale dataset of realistic images. We demonstrate the effectiveness of our realism reward function and the RL finetuning stage and show that \method{} can produce more diverse and photo-realistic materials compared to state-of-the-art material generators.

\vspace{-0.07in}
\section*{Acknowledgments}
This project was funded in part by the NSF CAREER Award $\#2238193$ and a generous gift from Adobe.

\vspace{-0.07in}
\section*{Ethical Statement}

We conducted anonymous user studies with voluntary participation. No personally identifiable information was collected or reported and the results are presented only in aggregate. We are not aware of any malicious use cases or foreseeable negative societal consequences arising from our proposed method.

\bibliographystyle{eg-alpha-doi} 
\bibliography{references}

@article{vecchio2024controlmat,
  author = {Vecchio, Giuseppe and Martin, Rosalie and Roullier, Arthur and Kaiser, Adrien and Rouffet, Romain and Deschaintre, Valentin and Boubekeur, Tamy},
  title = {ControlMat: A Controlled Generative Approach to Material Capture},
  year = {2024},
  issue_date = {October 2024},
  publisher = {Association for Computing Machinery},
  address = {New York, NY, USA},
  volume = {43},
  number = {5},
  issn = {0730-0301},
  url = {https://doi.org/10.1145/3688830},
  doi = {10.1145/3688830},
  journal = {ACM Trans. Graph.},
  month = {sep},
  articleno = {164},
  numpages = {17}
}

@article{Guarnera2016,
	journal = {Computer Graphics Forum},
	title = {BRDF Representation and Acquisition},
	author = {Guarnera, Dar'ya and Guarnera, Giuseppe Claudio and Ghosh, Abhijeet and Denk, Cornelia and Glencross, Mashhuda},
	year = {2016},
	publisher = {The Eurographics Association and John Wiley & Sons Ltd.},
}

@article{Li2017,
	author = {Li, Xiao and Dong, Yue and Peers, Pieter and Tong, Xin},
	title = {Modeling Surface Appearance from a Single Photograph Using Self-Augmented Convolutional Neural Networks},
	year = {2017},
	volume = {36},
	number = {4},
	journal = {ACM Trans. Graph.},
	pages = {45:1--45:11},
}

@article{Matusik2003,
  author = {Matusik, Wojciech and Pfister, Hanspeter and Brand, Matt and McMillan, Leonard},
  title = {A Data-Driven Reflectance Model},
  year = {2003},
  volume = {22},
  number = {3},
  journal = {ACM Trans. Graph.},
  pages = {759--769},
}

@article{zhou2022look,
  title={Look-Ahead Training with Learned Reflectance Loss for Single-Image SVBRDF Estimation},
  author={Zhou, Xilong and Kalantari, Nima Khademi},
  journal={ACM Transactions on Graphics (TOG)},
  volume={41},
  number={6},
  pages={1--12},
  year={2022},
  publisher={ACM New York, NY, USA}
}

@article{Deschaintre2018,
 author = {Deschaintre, Valentin and Aittala, Miika and Durand, Fredo and Drettakis, George and Bousseau, Adrien},
 title = {Single-image SVBRDF Capture with a Rendering-aware Deep Network},
 journal = {ACM Trans. Graph.},
 volume = {37},
 number = {4},
 year = {2018},
 pages = {128:1--128:15},
}

@inproceedings{Li2018,
  author = {Zhengqin Li and Kalyan Sunkavalli and Manmohan Chandraker},
  title     = {Materials for Masses: {SVBRDF} Acquisition with a Single Mobile Phone
               Image},
  booktitle = {Computer Vision - {ECCV} 2018 - 15th European Conference, Munich,
               Germany, September 8-14, 2018, Proceedings, Part {III}},
  series    = {Lecture Notes in Computer Science},
  volume    = {11207},
  pages     = {74--90},
  year      = {2018},
}

@inproceedings{zhou2022tilegen,
author = {Zhou, Xilong and Hasan, Milos and Deschaintre, Valentin and Guerrero, Paul and Sunkavalli, Kalyan and Kalantari, Nima Khademi},
title = {TileGen: Tileable, Controllable Material Generation and Capture},
year = {2022},
isbn = {9781450394703},
publisher = {Association for Computing Machinery},
address = {New York, NY, USA},
url = {https://doi.org/10.1145/3550469.3555403},
doi = {10.1145/3550469.3555403},
booktitle = {SIGGRAPH Asia 2022 Conference Papers},
articleno = {34},
numpages = {9},
location = {Daegu, Republic of Korea},
series = {SA '22}
}

@article{Martin2022,
author = {Martin, Rosalie and Roullier, Arthur and Rouffet, Romain and Kaiser, Adrien and Boubekeur, Tamy},
title = {MaterIA: Single Image High-Resolution Material Capture in the Wild},
journal = {Computer Graphics Forum},
volume = {41},
number = {2},
pages = {163-177},
doi = {https://doi.org/10.1111/cgf.14466},
url = {https://onlinelibrary.wiley.com/doi/abs/10.1111/cgf.14466},
year = {2022}
}

@misc{Substance,
	Author = {Adobe},
	Year = {2023},
	Note   = {Available at \url{https://substance3d.adobe.com/assets}. Accessed: 2026-02-03},
	Title = {Substance}
}

@article{Gao2019,
	title={Deep inverse rendering for high-resolution SVBRDF estimation from an arbitrary number of images},
	author={Gao, Duan and Li, Xiao and Dong, Yue and Peers, Pieter and Xu, Kun and Tong, Xin},
	journal={ACM Trans. Graph.},
	volume={38},
	number={4},
	year={2019}
}

@article{Deschaintre2019,
  author       = "Deschaintre, Valentin and Aittala, Miika and Durand, Fr\'edo and Drettakis, George and Bousseau, Adrien",
  title        = "Flexible SVBRDF Capture with a Multi-Image Deep Network",
  journal      = "Computer Graphics Forum",
  number       = "4",
  volume       = "38",
  year         = "2019",
}

@article{hu2022,
author = {Hu, Yiwei and He, Chengan and Deschaintre, Valentin and Dorsey, Julie and Rushmeier, Holly},
title = {An Inverse Procedural Modeling Pipeline for SVBRDF Maps},
year = {2022},
issue_date = {April 2022},
publisher = {Association for Computing Machinery},
address = {New York, NY, USA},
volume = {41},
number = {2},
issn = {0730-0301},
url = {https://doi.org/10.1145/3502431},
doi = {10.1145/3502431},
journal = {ACM Trans. Graph.},
month = {jan},
articleno = {18},
numpages = {17}
}

@article{Guo2020,
  title={MaterialGAN: Reflectance Capture using a Generative SVBRDF Model},
  author={Guo, Yu and Smith, Cameron and Ha\v{s}an, Milo\v{s} and Sunkavalli, Kalyan and Zhao, Shuang},
  journal={ACM Trans. Graph.},
  volume={39},
  number={6},
  year={2020},
  pages={254:1--254:13}
}

@article {Zhou2021,
journal = {Computer Graphics Forum},
title = {{Adversarial Single-Image SVBRDF Estimation with Hybrid Training}},
author = {Zhou, Xilong and Kalantari, Nima Khademi},
year = {2021},
publisher = {The Eurographics Association and John Wiley & Sons Ltd.},
ISSN = {1467-8659},
DOI = {10.1111/cgf.142635}
}

@article{Guo2021,
author = {Guo, Jie and Lai, Shuichang and Tao, Chengzhi and Cai, Yuelong and Wang, Lei and Guo, Yanwen and Yan, Ling-Qi},
title = {Highlight-Aware Two-Stream Network for Single-Image SVBRDF Acquisition},
year = {2021},
issue_date = {August 2021},
publisher = {Association for Computing Machinery},
address = {New York, NY, USA},
volume = {40},
number = {4},
issn = {0730-0301},
url = {https://doi.org/10.1145/3450626.3459854},
doi = {10.1145/3450626.3459854},
journal = {ACM Trans. Graph.},
month = {jul},
articleno = {123},
numpages = {14}
}

@inproceedings{GAN,
  title = {Generative Adversarial Nets},
  author = {Goodfellow, Ian and Pouget-Abadie, Jean and Mirza, Mehdi and Xu, Bing and Warde-Farley, David and Ozair, Sherjil and Courville, Aaron and Bengio, Yoshua},
  booktitle = {Advances in Neural Information Processing Systems 27},
  pages = {2672--2680},
  year = {2014},
}

@article{Henzler2021,
author = {Henzler, Philipp and Deschaintre, Valentin and Mitra, Niloy J. and Ritschel, Tobias},
title = {Generative Modelling of BRDF Textures from Flash Images},
year = {2021},
issue_date = {December 2021},
publisher = {Association for Computing Machinery},
address = {New York, NY, USA},
volume = {40},
number = {6},
journal = {ACM Trans. Graph.},
month = {dec},
articleno = {284},
numpages = {13},
}

@article{guerrero2022matformer,
    author = {Guerrero, Paul and Hasan, Milos and Sunkavalli, Kalyan and Mech, Radomir and Boubekeur, Tamy and Mitra, Niloy},
    title = {MatFormer: A Generative Model for Procedural Materials},
    year = {2022},
    volume = {41},
    number = {4},
    doi = {10.1145/3528223.3530173},
    journal = {ACM Trans. Graph.},
    articleno = {46}
}

@inproceedings{vecchio2024matfuse,
  title={Matfuse: controllable material generation with diffusion models},
  author={Vecchio, Giuseppe and Sortino, Renato and Palazzo, Simone and Spampinato, Concetto},
  booktitle={Proceedings of the IEEE/CVF Conference on Computer Vision and Pattern Recognition},
  pages={4429--4438},
  year={2024}
}

@inproceedings{xue2024reflectancefusion,
  title={ReflectanceFusion: Diffusion-based text to SVBRDF Generation},
  author={Xue, Bowen and Guarnera, Claudio and Zhao, Shuang and Montazeri, Zahra},
  booktitle={Eurographics Symposium on Rendering},
  year={2024},
  organization={Eurographics Association}
}

@inproceedings{zhou2023photomat,
  title={Photomat: A material generator learned from single flash photos},
  author={Zhou, Xilong and Hasan, Milos and Deschaintre, Valentin and Guerrero, Paul and Hold-Geoffroy, Yannick and Sunkavalli, Kalyan and Kalantari, Nima Khademi},
  booktitle={ACM SIGGRAPH 2023 Conference Proceedings},
  pages={1--11},
  year={2023}
}

@inproceedings{lin2024common,
  title={Common diffusion noise schedules and sample steps are flawed},
  author={Lin, Shanchuan and Liu, Bingchen and Li, Jiashi and Yang, Xiao},
  booktitle={Proceedings of the IEEE/CVF winter conference on applications of computer vision},
  pages={5404--5411},
  year={2024}
}

@inproceedings{rombach2022high,
  title={High-resolution image synthesis with latent diffusion models},
  author={Rombach, Robin and Blattmann, Andreas and Lorenz, Dominik and Esser, Patrick and Ommer, Bj{\"o}rn},
  booktitle={Proceedings of the IEEE/CVF conference on computer vision and pattern recognition},
  pages={10684--10695},
  year={2022}
}

@inproceedings{podell2023sdxl,
    title={{SDXL}: Improving Latent Diffusion Models for High-Resolution Image Synthesis},
    author={Dustin Podell and Zion English and Kyle Lacey and Andreas Blattmann and Tim Dockhorn and Jonas M{\"u}ller and Joe Penna and Robin Rombach},
    booktitle={The Twelfth International Conference on Learning Representations},
    year={2024},
    url={https://openreview.net/forum?id=di52zR8xgf}
}

@article{schuhmann2022laion,
  title={Laion-5b: An open large-scale dataset for training next generation image-text models},
  author={Schuhmann, Christoph and Beaumont, Romain and Vencu, Richard and Gordon, Cade and Wightman, Ross and Cherti, Mehdi and Coombes, Theo and Katta, Aarush and Mullis, Clayton and Wortsman, Mitchell and others},
  journal={Advances in Neural Information Processing Systems},
  volume={35},
  pages={25278--25294},
  year={2022}
}

@inproceedings{
    ddpo,
    title={Training Diffusion Models with Reinforcement Learning},
    author={Kevin Black and Michael Janner and Yilun Du and Ilya Kostrikov and Sergey Levine},
    booktitle={The Twelfth International Conference on Learning Representations},
    year={2024},
    url={https://openreview.net/forum?id=YCWjhGrJFD}
}

@inproceedings{li2023instant3d,
    title={Instant3D: Fast Text-to-3D with Sparse-view Generation and Large Reconstruction Model},
    author={Jiahao Li and Hao Tan and Kai Zhang and Zexiang Xu and Fujun Luan and Yinghao Xu and Yicong Hong and Kalyan Sunkavalli and Greg Shakhnarovich and Sai Bi},
    booktitle={The Twelfth International Conference on Learning Representations},
    year={2024},
    url={https://openreview.net/forum?id=2lDQLiH1W4}
}

@misc{schulman2017proximal,
      title={Proximal Policy Optimization Algorithms}, 
      author={John Schulman and Filip Wolski and Prafulla Dhariwal and Alec Radford and Oleg Klimov},
      year={2017},
      eprint={1707.06347},
      archivePrefix={arXiv},
      primaryClass={cs.LG},
      url={https://arxiv.org/abs/1707.06347}, 
}

@inproceedings{xie2024carve3d,
  title={Carve3d: Improving multi-view reconstruction consistency for diffusion models with rl finetuning},
  author={Xie, Desai and Li, Jiahao and Tan, Hao and Sun, Xin and Shu, Zhixin and Zhou, Yi and Bi, Sai and Pirk, S{\"o}ren and Kaufman, Arie E},
  booktitle={Proceedings of the IEEE/CVF Conference on Computer Vision and Pattern Recognition},
  pages={6369--6379},
  year={2024}
}

@inproceedings{loshchilov2017decoupled,
    title={Decoupled Weight Decay Regularization},
    author={Ilya Loshchilov and Frank Hutter},
    booktitle={International Conference on Learning Representations},
    year={2019},
    url={https://openreview.net/forum?id=Bkg6RiCqY7},
}

@inproceedings{radford2021learning,
  title={Learning transferable visual models from natural language supervision},
  author={Radford, Alec and Kim, Jong Wook and Hallacy, Chris and Ramesh, Aditya and Goh, Gabriel and Agarwal, Sandhini and Sastry, Girish and Askell, Amanda and Mishkin, Pamela and Clark, Jack and others},
  booktitle={International conference on machine learning},
  pages={8748--8763},
  year={2021},
  organization={PMLR}
}

@article{ho2020denoising,
  title={Denoising diffusion probabilistic models},
  author={Ho, Jonathan and Jain, Ajay and Abbeel, Pieter},
  journal={Advances in neural information processing systems},
  volume={33},
  pages={6840--6851},
  year={2020}
}

@inproceedings{luo2024single,
  title={Single-Image SVBRDF Estimation with Learned Gradient Descent},
  author={Luo, Xuejiao and Scandolo, Leonardo and Bousseau, Adrien and Eisemann, Elmar},
  booktitle={Computer Graphics Forum},
  volume={43},
  number={2},
  pages={e15018},
  year={2024},
  organization={Wiley Online Library}
}

@article{nie2024single,
  title={Single-Image SVBRDF Estimation Using Auxiliary Renderings as Intermediate Targets},
  author={Nie, Yongwei and Yu, Jiaqi and Long, Chengjiang and Zhang, Qing and Li, Guiqing and Cai, Hongmin},
  journal={IEEE Transactions on Visualization and Computer Graphics},
  year={2024},
  publisher={IEEE}
}

@inproceedings{zhou2023semi,
  title={A Semi-Procedural Convolutional Material Prior},
  author={Zhou, Xilong and Ha{\v{s}}an, Milo{\v{s}} and Deschaintre, Valentin and Guerrero, Paul and Sunkavalli, Kalyan and Kalantari, Nima Khademi},
  booktitle={Computer Graphics Forum},
  volume={42},
  number={6},
  pages={e14781},
  year={2023},
  organization={Wiley Online Library}
}

@inproceedings{wang2023deepbasis,
  title={DeepBasis: Hand-Held Single-Image SVBRDF Capture via Two-Level Basis Material Model},
  author={Wang, Li and Zhang, Lianghao and Gao, Fangzhou and Zhang, Jiawan},
  booktitle={SIGGRAPH Asia 2023 Conference Papers},
  pages={1--11},
  year={2023}
}

@inproceedings{zhang2023deep,
  title={Deep SVBRDF Estimation from Single Image under Learned Planar Lighting},
  author={Zhang, Lianghao and Gao, Fangzhou and Wang, Li and Yu, Minjing and Cheng, Jiamin and Zhang, Jiawan},
  booktitle={ACM SIGGRAPH 2023 Conference Proceedings},
  pages={1--11},
  year={2023}
}

@inproceedings{sartor2023matfusion,
  title={Matfusion: a generative diffusion model for svbrdf capture},
  author={Sartor, Sam and Peers, Pieter},
  booktitle={SIGGRAPH Asia 2023 Conference Papers},
  pages={1--10},
  year={2023}
}

@inproceedings{hu2023generating,
  title={Generating Procedural Materials from Text or Image Prompts},
  author={Hu, Yiwei and Guerrero, Paul and Hasan, Milos and Rushmeier, Holly and Deschaintre, Valentin},
  booktitle={ACM SIGGRAPH 2023 Conference Proceedings},
  pages={1--11},
  year={2023}
}

@inproceedings{he2023text2mat,
    booktitle = {Pacific Graphics Short Papers and Posters},
    editor = {Chaine, Raphaëlle and Deng, Zhigang and Kim, Min H.},
    title = {{Text2Mat: Generating Materials from Text}},
    author = {He, Zhen and Guo, Jie and Zhang, Yan and Tu, Qinghao and Chen, Mufan and Guo, Yanwen and Wang, Pengyu and Dai, Wei},
    year = {2023},
    publisher = {The Eurographics Association},
    ISBN = {978-3-03868-234-9},
    DOI = {10.2312/pg.20231275}
}

@article{johnson2019billion,
  title={Billion-scale similarity search with {GPUs}},
  author={Johnson, Jeff and Douze, Matthijs and J{\'e}gou, Herv{\'e}},
  journal={IEEE Transactions on Big Data},
  volume={7},
  number={3},
  pages={535--547},
  year={2019},
  publisher={IEEE}
}

@inproceedings{kingma2014,
  title     = {Auto-Encoding Variational Bayes},
  author    = {Kingma, Diederik P. and Welling, Max},
  booktitle = {International Conference on Learning Representations},
  year      = {2014}
}

@inproceedings{Yan:2023:PSDR-Room,
  title={PSDR-Room: Single Photo to Scene using Differentiable Rendering},
  author={Yan, K. and Luan, F. and Ha\v{s}an, M. and Groueix, T. and Deschaintre, V. and Zhao, S.},
  booktitle = {ACM SIGGRAPH Asia 2023 Conference Proceedings},
  year = {2023},
}

@InProceedings{zeroSNR,
    author    = {Lin, Shanchuan and Liu, Bingchen and Li, Jiashi and Yang, Xiao},
    title     = {Common Diffusion Noise Schedules and Sample Steps Are Flawed},
    booktitle = {Proceedings of the IEEE/CVF Winter Conference on Applications of Computer Vision (WACV)},
    month     = {January},
    year      = {2024},
    pages     = {5404-5411}
}

@inproceedings{
    hu2021lora,
    title={Lo{RA}: Low-Rank Adaptation of Large Language Models},
    author={Edward J Hu and Yelong Shen and Phillip Wallis and Zeyuan Allen-Zhu and Yuanzhi Li and Shean Wang and Lu Wang and Weizhu Chen},
    booktitle={International Conference on Learning Representations},
    year={2022},
    url={https://openreview.net/forum?id=nZeVKeeFYf9}
}

@software{Mitsuba3,
    title = {Mitsuba 3 renderer},
    author = {Wenzel Jakob and Sébastien Speierer and Nicolas Roussel and Merlin Nimier-David and Delio Vicini and Tizian Zeltner and Baptiste Nicolet and Miguel Crespo and Vincent Leroy and Ziyi Zhang},
    note   = {Available at \url{https://mitsuba-renderer.org}. Accessed: 2026-02-03},
    version = {3.1.1},
    year = 2022
}

@misc{envHDR,
  title = {PolyHaven HDR environments},
  author = {{Poly Haven}},
  year = {2024},
  note   = {Available at \url{https://polyhaven.com/hdris}. Accessed: 2026-02-03},
}

@inproceedings{johnson2016perceptual,
  title={Perceptual losses for real-time style transfer and super-resolution},
  author={Johnson, Justin and Alahi, Alexandre and Fei-Fei, Li},
  booktitle={Computer Vision--ECCV 2016: 14th European Conference, Amsterdam, The Netherlands, October 11-14, 2016, Proceedings, Part II 14},
  pages={694--711},
  year={2016},
  organization={Springer}
}

@misc{lee2023aligning,
      title={Aligning Text-to-Image Models using Human Feedback}, 
      author={Kimin Lee and Hao Liu and Moonkyung Ryu and Olivia Watkins and Yuqing Du and Craig Boutilier and Pieter Abbeel and Mohammad Ghavamzadeh and Shixiang Shane Gu},
      year={2023},
      eprint={2302.12192},
      archivePrefix={arXiv},
      primaryClass={cs.LG},
      url={https://arxiv.org/abs/2302.12192}, 
}

@article{dpok,
  title={Reinforcement learning for fine-tuning text-to-image diffusion models},
  author={Fan, Ying and Watkins, Olivia and Du, Yuqing and Liu, Hao and Ryu, Moonkyung and Boutilier, Craig and Abbeel, Pieter and Ghavamzadeh, Mohammad and Lee, Kangwook and Lee, Kimin},
  journal={Advances in Neural Information Processing Systems},
  volume={36},
  year={2024}
}

@inproceedings{vecchio2024matsynth,
  author    = {Vecchio, Giuseppe and Deschaintre, Valentin},
  title     = {MatSynth: A Modern PBR Materials Dataset},
  booktitle = {Proceedings of the IEEE/CVF Conference on Computer Vision and Pattern Recognition (CVPR)},
  month     = {June},
  year      = {2024},
  pages     = {22109-22118}
}

@misc{prabhudesai2023aligning,
      title={Video Diffusion Alignment via Reward Gradients}, 
      author={Mihir Prabhudesai and Russell Mendonca and Zheyang Qin and Katerina Fragkiadaki and Deepak Pathak},
      year={2024},
      eprint={2407.08737},
      archivePrefix={arXiv},
      primaryClass={cs.CV},
      url={https://arxiv.org/abs/2407.08737}, 
}

@article{ma2023opensvbrdf,
  title={OpenSVBRDF: A database of measured spatially-varying reflectance},
  author={Ma, Xiaohe and Xu, Xianmin and Zhang, Leyao and Zhou, Kun and Wu, Hongzhi},
  journal={ACM Transactions on Graphics (TOG)},
  volume={42},
  number={6},
  pages={1--14},
  year={2023},
  publisher={ACM New York, NY, USA}
}

@inproceedings{
clark2023directly,
title={Directly Fine-Tuning Diffusion Models on Differentiable Rewards},
author={Kevin Clark and Paul Vicol and Kevin Swersky and David J. Fleet},
booktitle={The Twelfth International Conference on Learning Representations},
year={2024},
url={https://openreview.net/forum?id=1vmSEVL19f}
}

@article{xu2023imagereward,
  title={Imagereward: Learning and evaluating human preferences for text-to-image generation},
  author={Xu, Jiazheng and Liu, Xiao and Wu, Yuchen and Tong, Yuxuan and Li, Qinkai and Ding, Ming and Tang, Jie and Dong, Yuxiao},
  journal={Advances in Neural Information Processing Systems},
  volume={36},
  pages={15903--15935},
  year={2023}
}

@inproceedings{wu2024deep,
  title={Deep reward supervisions for tuning text-to-image diffusion models},
  author={Wu, Xiaoshi and Hao, Yiming and Zhang, Manyuan and Sun, Keqiang and Huang, Zhaoyang and Song, Guanglu and Liu, Yu and Li, Hongsheng},
  booktitle={European Conference on Computer Vision},
  pages={108--124},
  year={2024},
  organization={Springer}
}

\include{sec/figures}

\end{document}